\documentclass[final,3p,times]{elsarticle}

\usepackage{color}

\newcommand{\be}{\begin{equation}}
\newcommand{\ee}{\end{equation}}

\newcommand{\la}{\langle}
\newcommand{\ra}{\rangle}

\newcommand{\e}{{\rm{e}}}

\newcommand{\flambda}{{\widehat{\lambda}}}
\newcommand{\lpsi}{{\widetilde{\psi}}}

\newcommand{\br}{{\bf{r}}}
\newcommand{\bv}{{\bf{v}}}

\usepackage{amssymb}

\journal{Physica A}

\begin{document}

\begin{frontmatter}



\title{Short note on the emergence of fractional kinetics\footnote{The present paper is published in Physica A 409 (2014) 29--34}}


\author{Gianni PAGNINI}
\ead{gpagnini@bcamath.org}
\address{BCAM - Basque Center for Applied Mathematics,\\
Alameda de Mazarredo 14, E--48009 Bilbao, Basque Country -- Spain \\
and\\
Ikerbasque, Basque Foundation for Science,\\
Alameda Urquijo 36-5, Plaza Bizkaia, E--48011 Bilbao, Basque Country -- Spain
}

\begin{abstract}
In the present Short Note an idea is proposed to  
explain the emergence and the observation of processes in complex media that are driven by
fractional non-Markovian master equations.
Particle trajectories are assumed to be solely Markovian and described by the Continuous Time Random Walk model.
But, as a consequence of the complexity of the medium,
each trajectory is supposed to scale in time according to a particular random timescale.
The link from this framework to microscopic dynamics is discussed and 
the distribution of timescales is computed.
In particular, when a stationary distribution is considered,
the timescale distribution is uniquely determined as a function related to the fundamental solution 
of the space-time fractional diffusion equation.
In contrast, when the non-stationary case is considered, the timescale distribution is no longer unique.
Two distributions are here computed: 
one related to the M-Wright/Mainardi function, which is Green's function of the time-fractional diffusion equation, 
and another related to the Mittag--Leffler function, which is the solution of the fractional-relaxation equation. 
\end{abstract}


\end{frontmatter}


\section{Introduction}
\label{sec:introduction}
Fractional kinetics is associated to phenomena governed by equations built on fractional derivatives.
This approach has turned out to be successful in modeling {\it anomalous diffusion} processes.

The label {\it anomalous diffusion} is used in contrast to {\it normal diffusion},
where the adjective {\it normal} has the double aim of highlighting that a Gaussian based process
is considered (because of the correspondence between the Normal and the Gaussian density)
and that it is a typical and usual diffusion process.
The observation in nature of anomalous diffusion has been definitively established,
see e.g. \cite{mercadier_etal-np-2009,ratynskaia_etal-prl-2006,barkai_etal-pt-2012}.

A number of stochastic approaches to explaining anomalous diffusion has been introduced in the literature.
One of the most successful is the Continuous Time Random Walk (CTRW)
\cite{fulger_etal-pre-2008, gorenflo_etal-jcam-2009,
hilfer_etal-pre-1995,klafter_etal-pra-1987,mainardi_etal-pa-2000,montroll_etal-jmp-1965,scalas_etal-pre-2004}.

However, 
recalling the simplicity of physical laws, 
in the {\it Simple Lessons from Complexity} taught by Goldenfeld \& Kadanoff \cite{goldenfeld_etal-s-1999},
the authors' first reply to the question
{\it ''So why, if the laws are so simple, is the world so complicated?''} is

\smallskip
{\it ''To us, complexity means that we have structure with variations.
Thus a living organism is complex because it has many different working parts, each formed by variations in the working out
of the same genetic code''}

\smallskip
\noindent
and finally that

\smallskip
{\it ''Complex systems form structures, and these structures vary widely in size and duration.
Their probability distributions are rarely normal, so that exceptional events are not rare.''}

\smallskip
With this in mind, here solely the simplest CTRW model is considered, i.e. the Markovian one.
However, notwithstanding this simplest framework, it is argued that anomalous diffusion emerges as a consequence of the
underlying variations of the structures of the medium from which a {\it wide} range of random timescales follows.
Hence, each particle trajectory is supposed to scale in time according to its own timescale.
This because any trajectory realization is supposed to occur in a random configuration of the medium
characterized by its own timescale.
In other words, the main idea discussed in this Short Note is that processes can be in general simply Markovian,
but, during the observation procedure, what is actually measured is the superposition of processes
of the same type but with different reference scales, as a consequence of structure variations.

Randomness of the timescale can be re-phrased as fluctuations of the timescale. 
This latter concept can be linked with the pioneering work by Beck \cite{beck-prl-2001} that has led to so-called 
superstatistics \cite{beck_etal-pa-2003}. 
However, the present research differs from that because it starts from a different explanatory idea and
it is based on particle trajectories modeled by a CTRW rather than by the Langevin equation, even though in both cases a superposition integral is used.
Moreover, through the velocity autocorrelation function, it has a connection with the microscopic dynamics
described by a Hamiltonian approach for the Brownian motion \cite{west_etal-2003} 
and by the so-called ``semidynamical'' {\it V-Langevin} approach \cite{balescu-2005,balescu-csf-2007}.

The present research is quite close to a recent work by Pramukkul {\it et al.} \cite{pramukkul_etal-amp-2013}.
In particular, since the CTRW is adopted, the present formalism can be related to the so-called 
stochastic central limit theorem \cite{pramukkul_etal-amp-2013}.
But, again, in the present research, the superposition formalism is introduced 
by some arguments linked to microscopic dynamics and not as a mathematical tool.
Moreover, with respect to the work by Pramukkul {\it et al.} \cite{pramukkul_etal-amp-2013},
here some results concerning the timescale distribution function are also presented and discussed.

The main features of CTRW are the following.
Let $p(\br,t)$ be the $pdf$ for a particle to be at $\br$ at the time $t$.
Moreover, let $\lambda(\delta \br)$ be the $pdf$ for a particle to make a jump of length $\delta \br$
after a waiting time $\tau$ whose $pdf$ is denoted by $\psi(\tau)$.
Since the integral $\displaystyle{\int_0^\tau \psi(\xi) \, d\xi}$ represents the probability that at least
one step is made in the temporal interval ($0,\tau$) \cite{mainardi_etal-pa-2000,scalas_etal-pa-2000},
the probability that a given waiting interval between two consecutive jumps is greater than or equal to $\tau$ is
$\displaystyle{\Psi(\tau)=1-\int_0^\tau \psi(\xi) \, d\xi}$ and the equation 
\be
\psi(\tau)=-\frac{d \Psi}{d \tau} 
\label{psieq}
\ee
holds \cite{mainardi_etal-pa-2000,scalas_etal-pa-2000}.
Hence $\Psi(t)$ is the probability that, after a jump, the diffusing quantity does not change during
the temporal interval of duration $\tau$ and
it is the {\it survival probability} at the initial position \cite{hilfer_etal-pre-1995}.

When jumps and waiting times are statistically independent,
the master equation of the CTRW model is \cite{mainardi_etal-pa-2000}
\be
\int_0^t \Phi(t-\tau) \frac{\partial p}{\partial \tau} \, d\tau=-p(\br,t) + \sum_{\br'} \lambda(\br-\br')p(\br',t) \,,
\label{ctrwmainardi}
\ee
where $\widetilde{\Phi}(s)=\widetilde{\Psi}(s)/\lpsi(s)$ is a memory kernel and tilde symbol $\, \widetilde{\cdot} \,$ means the Laplace transform. 

From equation (\ref{ctrwmainardi}) it follows that a Markovian process
is obtained when $\Phi(\tau)=\delta(\tau)$,
which implies that $\widetilde{\Phi}(s)=1$ so that 
$\widetilde{\Psi}(s)=\lpsi(s)$ and finally $\Psi(\tau)=\psi(\tau)$.
Functions $\Psi(\tau)$ and $\psi(\tau)$ are related by (\ref{psieq}), so then a CTRW model is Markovian if
$\Psi(\tau)=\e^{-\tau}$.
Equivalently, when $\Psi(\tau)$ is different from an exponential function, the resulting CTRW model is non-Markovian
or, by using mathematical terminology, it belongs to the class of semi-Markov process.

Assume a complex medium is formed by randomly variable structures characterized by individual scales for each configuration.
Hence, for any Markovian-CTRW trajectory,
the time variable $t$ and the waiting-time $\tau$ have to be scaled by a particular random timescale $T$.
In particular, the survival probability $\Psi(\tau)$ turns out to be
\be
\Psi(\tau)=\Psi(\tau/T)=\e^{-\tau/T} \,.
\label{Psiexp}
\ee
The ratio $\tau/T$ for any observation time is a random variable because $T$ is a random variable.

In a pioneering paper by Hilfer \& Anton in 1995 \cite{hilfer_etal-pre-1995}, it was shown that
if the survival probability $\Psi(\tau)$ is 
\be
\Psi(\tau)=E_\beta(-\tau^\beta) \,, \quad 0<\beta<1 \,,
\label{PsiML}
\ee 
where $E_\beta(z)$ is the Mittag--Leffler function defined as 
\cite[Appendix E]{haubold_etal-jam-2011,mainardi-2010}
\be
E_\beta(z)=\sum_{n=0}^\infty \frac{z^n}{\Gamma(\beta n+1)} \,, \quad z \in C \,,
\label{ML}
\ee
then the process is governed by the following fractional non-Markovian master equation
\be
\frac{\partial^\beta p}{\partial t^\beta}=-p(\br,t) + \sum_{\br'} \lambda(\br-\br') p(\br',t) \,, \quad 0 < \beta < 1 \,,
\label{fcctrw}
\ee
where $\displaystyle{\frac{\partial^\beta}{\partial t^\beta}}$ can be the fractional derivative operator
both in the Riemann--Liouville and in the Caputo sense \cite{gorenflo_etal-cism-1997}. 
It is well known that a survival probability of the Mittag--Leffler type (\ref{PsiML}), when $0<\beta<1$, 
decreases asymptotically for $\tau \to \infty$ with the power law $\tau^{-\beta}$, see e.g.
\cite[Section 1.3]{mainardi-2010}, 
thus slower than the survival probability of exponential type (\ref{Psiexp})
and so large-waiting-time events are not rare,
in agreement with Goldenfeld \& Kadanoff \cite{goldenfeld_etal-s-1999}.
The Markovian case is recovered when $\beta=1$, because $E_1(-z)=\e^{-z}$.
Non-Markovian CTRW models with a survival probability of the Mittag--Leffler type (\ref{PsiML})
have been widely studied,
see e.g. \cite{fulger_etal-pre-2008,gorenflo_etal-jcam-2009,scalas_etal-pre-2004}.

Let $f(T,t)$ be the probability density function of timescale $T$ 
with normalization condition $\displaystyle{\int_0^\infty f(T,t) \, dT=1}$.
Then a non-Markovian process follows from any $f(T,t)$ for which 
\be
\int_0^\infty \Psi_{\! M}(t/T) \, f(T,t) \, dT=\Psi(t) \,, \quad \Psi_{\! M}(t/T)=\e^{-t/T} 
\label{superpositiongen}
\ee
holds, where $M$ stands for Markovian.
In particular, when (\ref{PsiML}) holds, formula (\ref{superpositiongen}) reads
\be
\int_0^\infty \e^{-t/T} \, f(T,t) \, dT = E_\beta(-t^\beta) \,,
\quad 0 < \beta < 1 \,,
\label{superposition}
\ee
and a fractional kinetics governed by non-Markovian master equation (\ref{fcctrw}) arises.

Equations (\ref{superposition}) and (\ref{superpositiongen}) embody the relationship between the present approach
and the stochastic central limit theorem, see \cite[Equation (36)]{pramukkul_etal-amp-2013}.

However, the CTRW model is a purely random process and any relationships with the microscopic dynamical laws 
are neglected. The microscopic level is generally described by adopting the deterministic Hamiltonian formalism 
of classical mechanics.
But the Hamiltonian approach is very often highly difficult and prohibitive for handling transport phenomena.
The hard task of mapping Hamiltonian and dissipative systems is not 
attempted in the present Short Note.

Nevertheless, the derivation of the motion of a Brownian particle in a fluid of lighter particles
from a Hamiltonian description has been presented by West {\it et al.} \cite[Section 1.2.1]{west_etal-2003}.
The energy has been split into the sum of the fluid energy, 
the energy of the Brownian particle and the energy for their interaction.
In particular, the fluid background is modeled as a heat bath with a system of harmonic oscillators.
Furthermore a generalized Langevin equation is derived 
where the stochastic interpretation of the driving forcing is introduced by noting that the initial state of the bath is
uncertain and it is in general only determined via a distribution of initial states
\cite[Equation (1.34)]{west_etal-2003}.
Under certain conditions,
the velocity autocorrelation function $R(t)$ turns out to satisfy the integral equation
\cite[Equation (1.72)]{grigolini-acp-1985,grigolini_etal-pre-1999,west_etal-2003} 
\be
\frac{\partial R}{\partial t} = - \, C \, \int_0^t \kappa(t-\tau) R(\tau) \, d\tau \,,
\ee
where $\kappa(t)$ represents the correlation function of the stochastic force in the corresponding generalized Langevin equation
and $C$ is a particular constant related to the Hamiltonian function.

A further tractable starting point, that is closer than CTRW to the microscopic dynamics, 
is provided by a ``semidynamical'' formalism based on an equation of motion of Newtonian (or Hamiltonian) type for 
a tracer particle moving in the presence of a random potential.
This is an intermediate description in which some statistical properties of the microscopic particle velocity
are supposed to be known. It is called {\it V-Langevin} equation and it reads \cite{balescu-2005,balescu-csf-2007}
\be
\frac{d \br}{dt}=\bv(t) \,, \quad \la v_i(t) \ra =0 \,, \quad \la v_i(0)v_j(t) \ra = R_{ij}(t) \,.
\label{V-Langevin}
\ee
In the {\it one-dimensional} case, a realistic microscopic model 
has been derived by Balescu assuming the local approximation \cite{balescu-2005,balescu-csf-2007}.
The evolution equation for the particle $pdf$ turns out to be \cite{balescu-2005,balescu-csf-2007}
\be
\frac{\partial p}{\partial t}=\frac{\partial^2}{\partial r^2}\int_0^t R(t-\tau) \, p(r,\tau) \, d\tau \,.
\label{Balescu}
\ee
Equation (\ref{Balescu}) was derived also by Grigolini 
by using fundamental arguments of statistical mechanics and the condition that $v(t)$ is a 
two-state system \cite{grigolini-nnds-1989}; see also the derivation by Allegrini {\it et al.} \cite{allegrini_etal-pre-1996}. 

Through the autocorrelation function $R(t)$, what follows 
establishes a relation with the microscopic Hamiltonian description.
In fact, by applying the local approximation also to (\ref{ctrwmainardi}), i.e. 
$\flambda(\kappa)=1-\kappa^2 + \dots$ where symbol $\, \widehat{\cdot} \,$ means Fourier transform, 
its relationship with (\ref{Balescu}) becomes evident
and it holds \cite{balescu-csf-2007}
\be
\widetilde{\Psi}(s)=\frac{1}{s+\widetilde{R}(s)} \,.
\label{PsiR}
\ee
Indeed, the survival probability $\Psi(t)$ of a CTRW model is microscopically determined by the autocorrelation function $R(t)$
of the particle velocity.
From (\ref{superpositiongen}) it follows that even the distribution of timescale $f(T,t)$ is linked to  
the autocorrelation function $R(t)$ of microscopic velocity.
In fact
\begin{eqnarray}
\widetilde{\Psi}(s) 
&=& \int_0^\infty \e^{-st}\left\{\int_0^\infty \Psi_{\! M}(t/T) \, f(T,t) \, dT\right\} \, dt \nonumber \\
&=& \int_0^\infty \left\{\int_0^\infty \e^{-(s+1/T) \, t} \, f(T,t) \, dt\right\} \, dT 
= \int_0^\infty \widetilde{f}(T,s+1/T) \, dT \qquad
\end{eqnarray}  
and the distribution $f(T,t)$ is seen to be related to $\widetilde{R}(s)$ by
\be
\int_0^\infty \widetilde{f}(T,s+1/T) \, dT = \frac{1}{s + \widetilde{R}(s)} \,.
\label{fg}
\ee

To conclude the issue of microscopic connections, 
formula (\ref{PsiR}) due to Balescu \cite{balescu-csf-2007} establishes a connection between CTRW approach and microscopic
dynamics, while formula (\ref{fg}) establishes a connection between timescale distribution $f(T,t)$ and
microscopic dynamics.  
However, it is here recalled that formulae (\ref{PsiR}) and (\ref{fg}) 
hold solely in the {\it one-dimensional} case with local approximation,
but formula (\ref{superposition}) is of general validity.
Hence, formula (\ref{superposition}) is further analyzed.

In the framework of fractional kinetics,
when the survival probability $\Psi(t)$ is of the Mittag--Leffler type (\ref{PsiML}),
the equation $\widetilde{\Psi}(s)=s^{\beta-1}/(s^\beta+1)$ holds and from (\ref{PsiR}) it follows that $\widetilde{R}(s)=s^{1-\beta}$.
Let $T_M$ be the single existing timescale in the Markovian case,
then $\Psi(t/T_M)=\e^{-t/{T_M}}$ and $\widetilde{\Psi}(s)=1/(s+1/T_M)$ so that
from (\ref{PsiR}) $\widetilde{R}_M(s)=1/T_M$ and then $R_M(t)=\delta(t)/T_M$. 
Finally from (\ref{fg}) it follows that $f_M(T,t)=\delta(T-T_M)$ which leads to the single timescale $T=T_M$.

Distribution of timescales $f(T,t)$ in the non-Markovian case can be obtained by analysis of well-known results on the
Mittag--Leffler function (\ref{ML}). An updated list of results can be found in the book by Mainardi \cite[Appendix E]{mainardi-2010}
and in the review paper by Haubold {\it et al.} \cite{haubold_etal-jam-2011}. 

In particular, it is possible to establish uniquely the timescale distribution
in the {\it stationary} case, i.e. $f(T,t)=f^S(T)$.
In fact, it is well known that the equation
\be
\int_0^\infty \e^{-t y} \, K_\beta(y) \, dy= E_\beta(-t^\beta) \,, \quad 0 < \beta < 1 
\label{GM97}
\ee
holds \cite{gorenflo_etal-cism-1997,mainardi-2010}, 
where
\be
K_\beta(y)=\frac{1}{\pi} \frac{y^{\beta-1} \sin(\beta \pi)}{1+2 y^\beta \cos(\beta \pi) + y^{2\beta}} \,,
\label{Kbeta}
\ee
so that, comparing (\ref{superposition}) and (\ref{GM97}), it follows that the {\it stationary} timescale distribution $f^S(T)$ is
\be
f^S(T) = \frac{1}{T^2} K_\beta\left(\frac{1}{T}\right) \,.
\label{fstationary}
\ee

It is worth remarking that the distribution $K_\beta$ defined in (\ref{Kbeta}) is 
related to the fundamental solution
of the space-time fractional diffusion equation when both the space and time fractional orders of derivation are equal to $\beta$ 
and the asymmetry parameter assumes the extremal value such that the function support is the positive real axes 
\cite{mainardi_etal-fcaa-2001}. 
This $pdf$ has emerged in numerical studies on particle displacement in zonal flows
in non-diffusive chaotic transport by Rossby waves \cite{delcastillonegrete-npg-2010} and 
in plasma turbulence when Larmor effects are taken into account \cite{gustafson_etal-pp-2008}. 
Such a kind of fractional diffusion process is also referred 
to as neutral-fractional diffusion \cite{luchko-aipcp-2012}.
In the Markovian limit, i.e. $\beta=1$,
the relation $K_1(y)=\sin \pi/[\pi \, (y-1)^2] \to \delta(y-1)$ holds and a single timescale follows.

When a {\it non-stationary} distribution is desired, the timescale distribution $f(T,t)$ is no longer unique.
This can be shown by
applying the change of variable $t/T=q$ to the integral (\ref{superposition}) which becomes
\be
\int_0^\infty \e^{-q} \, \mathcal{H}(q,t) \, dq = E_\beta(-t^\beta) \,, \quad 0 < \beta < 1 \,,
\label{integraltransform}
\ee
where $\displaystyle{\mathcal{H}(q,t)=f(t/q,t) \, t/q^2}$ is the kernel of an integral transformation.
The same transformation pair $\e^{-q} \longleftrightarrow E_\beta(-t^\beta)$ can be obtained from different kernels.
In this respect, in what follows, two {\it non-stationary} distributions are derived.
Hence a further constraint is needed to select uniquely the timescale distribution in the {\it non-stationary} case.

A {\it non-stationary} distribution $f(T,t)$ can be obtained by the following formula \cite[Appendix F]{mainardi-2010}
\be
\int_0^\infty \e^{-s z} \, M_\beta(z) \, dz = E_\beta(-s) \,, \quad 0 < \beta < 1 \,,
\label{MML}
\ee
where
\be
M_\beta(z)=\sum_{n=0}^{\infty} \frac{(-z)^n}{n! \, \Gamma[-\beta n + (1-\beta)]} 
= \frac{1}{\pi} \, \sum_{n=1}^{\infty} \,\frac{(-z)^{n-1}}{(n-1)!} \, \Gamma(\beta n) \, \sin (\pi\beta n) \,,
\label{serieM}
\ee
so that, after the change of variables $zs =q$ and $s=t^\beta$, 
from comparing (\ref{integraltransform}) and (\ref{MML}) it follows that
\be
\mathcal{H}(q,t)= \frac{1}{t^\beta} M_\beta\left(\frac{q}{t^\beta}\right) \,,
\ee
and a {\it non-stationary} timescale distribution turns out to be
\be
f(T,t)=\frac{1}{T^2} \, t^{1-\beta} \, M_\beta\left(\frac{t^{1-\beta}}{T}\right) \,, \quad 0 < \beta < 1 \,.
\label{fM}
\ee
The Markovian case is recovered when $\beta=1$ because of the property $M_1(z)=\delta(z-1)$.  

The function $M_\beta$ is a transcendental function of the Wright type \cite{mainardi_etal-amc-2003,mainardi_etal-ijde-2010}.
It is also referred to as M-Wright/Mainardi function because it was originally obtained by Mainardi
as the fundamental solution of the time-fractional diffusion equation \cite{mainardi-csf-1996}.
It has been shown that when such fractional diffusion processes are properly characterized with stationary increments,
the M-Wright/Mainardi function plays the same key role as the Gaussian density for the standard and 
fractional Brownian motions \cite{mura_etal-jpa-2008,pagnini-fcaa-2013}.
The properties of the corresponding master equation lead to such diffusion processes being named as
Erd\'elyi--Kober fractional diffusion \cite{pagnini-epjst-2011,pagnini-fcaa-2012}.
Actually, the M-Wright/Mainardi function 
has turned out to be related also to
the quadratic variation for compound renewal processes \cite{scalas_etal-fcaa-2012}.
Further properties on the M-Wright/Mainardi function can be found in References 
\cite[Chapter 6 and Appendix F]{mainardi_etal-ijde-2010,mainardi-2010}.

A further {\it non-stationary} distribution can be derived by using the following formula \cite{mainardi-2010}
\be
\frac{1}{\sqrt{\pi \, t}} \int_0^\infty \e^{-z^2/(4 t)} \, E_{2 \beta}(-z^{2 \beta}) \, dz = E_\beta(-t^\beta) \,, 
\label{GaussML}
\ee
so that
\be
\mathcal{H}(q,t)=\frac{1}{4 \, \sqrt{\pi \, q} \, t} \, E_{2\beta}[-(4 \, q \, t)^\beta] \,,
\ee
and the corresponding {\it non-stationary} distribution is
\be
f(T,t)=\frac{1}{4 \, \sqrt{\pi \, t} \, T^{3/2}} \, E_{2\beta}\left[-\left(\frac{2 \, t}{T^{1/2}}\right)^{2\beta}\right] \,.
\label{fML}
\ee
In order to have a non-negative distribution, 
the parameter $\beta$ must be constrained by $0 < 2 \beta \le 1$, 
according to complete monotonicity property of the Mittag--Leffler function \cite[Appendix E]{mainardi-2010}.
This constraint avoids the possibility of recovering the Markovian limit $\beta=1$. 

It is worth highlighting that the
survival probability (\ref{PsiML}) and timescale distribution (\ref{fML}) are functions of the same type,
i.e. $E_\alpha(-z^\alpha)$ with $0 < \alpha < 1$,
which is the solution of the fractional-relaxation equation
\cite{mainardi-csf-1996,gorenflo_etal-cism-1997}. 
This strong relationship is in agreement with the physically sound idea that there exists a deep interrelation between 
the structure of the medium into which particles diffuse and the characteristics of particle jumps.
These correlated features of the system are embodied by
the timescale distribution $f(T,t)$ and by the survival probability $\Psi(t/T)$, respectively.

Summarizing, in this Short Note 
an idea is proposed to explain the emergence and observation of fractional kinetic processes.
The idea is based on the fact that any particle trajectory experiences a different random timescale 
as a consequence of the variations of the underlying structures.
Hence, 
observed processes can be understood as the superposition of particle trajectories driven by a simple Markovian CTRW model but,
as a consequence of the variations of the structures, each one scaled with its own timescale $T$.
This is in agreement with Goldenfeld \& Kadanoff's definition of complexity \cite{goldenfeld_etal-s-1999},
where complexity means that the medium forms structures that vary widely in size and duration.

In this formulation, a single real and measured time is considered and it is not turned into a random variable.
But, since 
any configuration of structures is assumed to be random, the characteristic timescale $T$ is random 
and distributed according to the probability density function $f(T,t)$.
The density $f(T,t)$ has emerged to be related to the autocorrelation function
of microscopic particle velocity.
However, the density $f(T,t)$ can be uniquely determined only when it is assumed to be {\it stationary}.

\section*{Acknowledgements}
The author would like to thank anonymous referees for useful remarks and  
also acknowledges helpful suggestions and support from Doctor P. Paradisi and Professors F. Mainardi, E. Scalas and D. Chillingworth. 

\bibliographystyle{elsarticle-num}
\bibliography{abbr,all}

\end{document}